\documentclass[12pt]{article}
\pdfoutput=1

\usepackage{fullpage}
\usepackage{amsmath, amssymb, hyperref}

\newcommand{\ud}{\text{d}}

\begin{document}

\title{A formal notion of genericity\\
       and term-by-term vanishing superpotentials\\
       at supersymmetric vacua\\
       from R-symmetric Wess-Zumino models}
\author{James Brister\textsuperscript{a, *},
        Zheng Sun\textsuperscript{a, \dag},
        Greg Yang\textsuperscript{b, \ddag}\\
        \textsuperscript{a}\normalsize\textit{College of Physics, Sichuan University,}\\
                           \normalsize\textit{29 Wangjiang Road, Chengdu, Sichuan, 610064, PRC}\\
        \textsuperscript{b}\normalsize\textit{Microsoft Research Lab – Redmond,}\\
                           \normalsize\textit{14820 NE 36th Street, Redmond, Washington, 98052, USA}\\
        \normalsize\textit{E-mail:}
        \textsuperscript{*}\texttt{jbrister@scu.edu.cn,}
        \textsuperscript{\dag}\texttt{sun\_ctp@scu.edu.cn,}
        \textsuperscript{\ddag}\texttt{gregyang@microsoft.com}
       }
\date{}
\maketitle

\begin{abstract}
It is known in previous literature that if a Wess-Zumino model with an R-symmetry gives a supersymmetric vacuum, the superpotential vanishes at the vacuum.  In this work, we establish a formal notion of genericity, and show that if the R-symmetric superpotential has generic coefficients, the superpotential vanishes term-by-term at a supersymmetric vacuum.  This result constrains the form of the superpotential which leads to a supersymmetric vacuum.  It may contribute to a refined classification of R-symmetric Wess-Zumino models, and find applications in string constructions of vacua with small superpotentials.  A similar result for a scalar potential system with a scaling symmetry is discussed.
\end{abstract}

\section{Introduction}

The relation between R-symmetries and supersymmetry (SUSY) breaking~\cite{Intriligator:2007cp} in generic Wess-Zumino models~\cite{Wess:1973kz, Wess:1974jb, ORaifeartaigh:1975nky} is described by the Nelson-Seiberg theorem~\cite{Nelson:1993nf} and its several recent extensions~\cite{Sun:2011fq, Kang:2012fn, Li:2020wdk, Brister:2021xxxx}, including counterexamples with generic parameters and special R-charge assignments~\cite{Sun:2019bnd, Amariti:2020lvx, Sun:2021svm, Li:2021ydn}.  This lore has manifested in the study of both SUSY phenomenology~\cite{Nilles:1983ge, Martin:1997ns, Baer:2006rs, Terning:2006bq, Dine:2007zp} and string phenomenology~\cite{Grana:2005jc, Douglas:2006es, Blumenhagen:2006ci, Ibanez:2012zz, Blumenhagen:2013fgp}.  The notion of genericity in these results usually refers to genericity of parameters, which is also related to the notion of naturalness.  A property, such as the existence of a SUSY vacuum, is a generic prediction of the model if it happens everywhere or in an open region of the parameter space.  The prediction is thus robust to small changes of parameters from a particular point in the parameter space, and no fine-tuning is needed.  Although non-generic R-charges are mentioned in counterexamples to the Nelson-Seiberg theorem, they are more properly described as special R-charge assignments satisfying certain conditions~\cite{Sun:2021svm}.  We therefore focus on generic parameters in this work.

In this work, we consider one possible formalization of genericity, and show that demanding that a property holds generically can lead to additional properties of the model.  In particular, we show that at a SUSY vacuum from a generic R-symmetric Wess-Zumino model, every term of the superpotential vanishes individually.  Thus each term of the superpotential must contain at least one field with a zero expectation value.  This result may contribute to a refined classification of R-symmetric Wess-Zumino models, and find applications in string constructions of vacua with small superpotentials.  The proof can be generalized to any generic function which takes a fixed value at a stationary point.  We discuss a scalar potential system with a scaling symmetry as an example.

The rest of this paper is arranged as follows.  Section \ref{sec:2} reviews the result that a SUSY vacuum from an R-symmetric Wess-Zumino model implies a vanishing superpotential.  Section \ref{sec:3} gives a formal notion of genericity used in this work.  Section \ref{sec:4} proves the result of a term-by-term vanishing superpotential at a SUSY vacuum from a generic R-symmetric Wess-Zumino model.  Section \ref{sec:5} discusses implications and generalizations of the result.

\section{SUSY vacua from R-symmetric models} \label{sec:2}

In a Wess-Zumino model, a SUSY vacuum corresponds to a solution to the equations
\begin{equation}
\partial_i W = \frac{\partial W}{\partial \phi_i}
             = 0,
\end{equation}
where $W = W(\phi_i)$ is a function named the superpotential for a set of scalar fields $\{ \phi_i \}$.  In the superspace formalism, $W$ is promoted to be a holomorphic function for a set of chiral superfields $\{ \Phi_i \}$, and appears in the SUSY action as
\begin{equation}
S_W = \int \ud^4 x \, \mathcal{L}_W
    = - \int \ud^4 x \,
             \left ( \int \ud \theta^\alpha \ud \theta_\alpha \, W + \text{c.c.} \right ).
\end{equation}
An R-symmetry acts non-trivially on the Grassmann numbers $\theta^\alpha$.  In $\mathcal{N}=1$ SUSY, a continuous R-symmetry is assumed to be $\text{U}(1)_\text{R}$.  The R-charge for $\theta^\alpha$ is conventionally set to $1$.  Thus the Berezin integral $\int \! \ud \theta^\alpha$ has R-charge $- 1$, and $W$ must have R-charge $2$ to make the SUSY action invariant under the R-symmetry.  Each field $\phi_i$ has R-charge $r_{(i)}$, thus the form of $W$ is constrained so that each term must be an R-charge $2$ combination of fields.  A finite transformation of $\text{U}(1)_\text{R}$ gives
\begin{equation}
\hat R(\zeta) \theta^\alpha = e^{i \zeta} \theta^\alpha, \quad
\hat R(\zeta) \phi_i = e^{i r_{(i)} \zeta} \phi_i, \quad
\hat R(\zeta) W = e^{2 i \zeta} W, \quad
\zeta \in \mathbb{R}.
\end{equation}
Taking an infinitesimal $\text{U}(1)_\text{R}$ transformation at $\zeta = 0$, we have
\begin{equation}
\left. \frac{\ud}{\ud \zeta} \right \rvert_0 \hat R(\zeta) W
    = 2 i W
    = \partial_i W \left. \frac{\ud}{\ud \zeta} \right \rvert_0 \hat R(\zeta) \phi_i
    = i r_{(i)} \phi_i \partial_i W.
\end{equation}
Thus the following identity is satisfied for any R-symmetric superpotential:
\begin{equation} \label{eq:2-01}
W = \frac{1}{2} r_{(i)} \phi_i \partial_i W.
\end{equation}
Using the Cauchy-Schwarz inequality, we obtain a bound on the superpotential~\cite{Dine:2009sw}:
\begin{equation} \label{eq:2-02}
\lvert \langle W \rangle \rvert
    = \frac{1}{2} \lvert \langle r_{(i)} \phi_i \partial_i W \rangle \rvert
    \le \frac{1}{2} \sqrt{r_{(i)}^2 \langle \phi_i^* \phi_i \rangle \langle (\partial_j W)^* \partial_j W \rangle}
    = \frac{1}{2} f_a f,
\end{equation}
where the R-axion decay constant $f_a$ and the Goldstino decay constant $f$ are defined as
\begin{align}
f_a &= \lVert r_{(i)} \langle \phi_i \rangle \rVert
     = \sqrt{r_{(i)}^2 \langle \phi_i^* \phi_i \rangle},\\
f &= \lVert \langle \partial_i W \rangle \rVert
   = \sqrt{\langle (\partial_i W)^* \partial_i W \rangle}.
\end{align}
At a SUSY vacuum, field expectation values satisfy $\partial_i W = 0$.  So we have $W = 0$ according to either Eq.~\eqref{eq:2-01} or Eq.~\eqref{eq:2-02}~\cite{Dine:2009sw, Kappl:2008ie}.

The result of a vanishing $W$ is general for any R-symmetric $W$ which gives a SUSY vacuum, and does not depend on whether $W$ is generic or not.  It is possible to arrange several nonzero terms which add up to zero.  For example, consider the superpotential
\begin{equation}
W = a X_1 + a X_2 + b X_1 Y^2 + b X_2 Y^2
  = (X_1 + X_2) (a + b Y^2)
\end{equation}
with the R-charge assignment
\begin{equation}
r_{(X_1)} = r_{(X_2)} = 2, \quad
r_{(Y)} = 0.
\end{equation}
It gives two infinite set of SUSY vacua at
\begin{equation}
X_1 = - X_2
    \in \mathbb{C}, \quad
Y = \pm \sqrt{- a / b}
\end{equation}
with degeneracy on a complex one-dimensional subspace of the $X_1$-$X_2$ space.  Generic nonzero values of $a$, $b$, $X_1$ and $X_2$ make all terms of $W$ nonzero, but these terms add up to zero, and $W = 0$ is satisfied at a SUSY vacuum.  However, the coefficients of $W$ are not independent, so $W$ is not generic in this example.  In the following sections, we are to show that such cancellation between terms does not happen in generic models:  every term of $W$ must vanish at a SUSY vacuum from a generic R-symmetric $W$.

\section{A formal notion of genericiy} \label{sec:3}

To establish a formal notion of genericity in a general model, we suppose that the model is described by a set of parameters $\{ c_\alpha \}$.  This description is applicable, but not limited, to the Wess-Zumino models in this work.  Consider a property represented as a function $F(c_\alpha)$ on the parameter space, such that $F(c_\alpha) = 0$ is satisfied if and only if the model with parameters at the point $c_\alpha$ possesses the relevant property.  The property is a generic prediction of the model if we have $F(c_\alpha) = 0$ in an open region near a certain parameter point $c_\alpha^{(0)}$, which is usually determined by experiments.  The property is therefore robust to a small perturbation of parameters from the point, and no fine-tuning is needed.

Most physics models have their dynamics described by a set of continuous variables $\{ \phi_i \}$, such as coordinates, wave functions or fields.  The equations of motion are expressed in these variables, and their solutions give the states and evolution of the physics system.  The property of our interest depends both explicitly on parameters $c_\alpha$, and on expectation values $\langle \phi_i \rangle = \phi_i(c_\alpha)$ which depend on $c_\alpha$.  A perturbation of $c_\alpha$ causes a shift of $\phi_i(c_\alpha)$.  The notion of genericity is then expressed as
\begin{equation} \label{eq:3-01}
F(\phi_i(c_\alpha^{(0)}), c_\alpha^{(0)}) = F(\phi_i(c_\alpha^{(0)} + \delta c_\alpha), c_\alpha^{(0)} + \delta c_\alpha)
                                          = 0, \quad
\forall \delta c_\alpha \text{ s.t.\ } \lvert \delta c_\alpha \rvert < \epsilon,
\end{equation}
where $\epsilon$ is the radius of a neighborhood of $c_\alpha^{(0)}$ in which the model retains the property.  Assuming the functions $F(\phi_i, c_\alpha)$ and $\phi_i(c_\alpha)$ are at least once differentiable at $c_\alpha^{(0)}$, which is true for most physics systems,  Eq.~\eqref{eq:3-01} then implies
\begin{equation} \label{eq:3-02}
\left. \frac{\ud F}{\ud c_\alpha} \right \lvert_{c_\beta^{(0)}}
    = \left. \left (\frac{\partial F}{\partial c_\alpha}
                    + \frac{\partial F}{\partial \phi_i}
                      \frac{\partial \phi_i}{\partial c_\alpha}
      \right ) \right \lvert_{c_\beta^{(0)}}
    = 0.
\end{equation}
Note that Eq.~\eqref{eq:3-02} is only a necessary condition for genericity.  In case that $F(\phi_i, c_\alpha)$ and $\phi_i(c_\alpha)$ are analytical functions at $c_\alpha^{(0)}$, the equivalent condition for Eq.~\eqref{eq:3-01} should be
\begin{equation} \label{eq:3-03}
\left. \frac{\ud^n F}{\ud c_\alpha^n} \right \lvert_{c_\beta^{(0)}} = 0, \quad
\forall n \in \mathbb{Z}_{\ge 0}.
\end{equation}
Just as happens in Eq.~\eqref{eq:3-02}, the total derivative in Eq.~\eqref{eq:3-03} can also be replaced by partial derivatives using the intermediate variables $\phi_i$.

Note that our notion of genericity takes the parameter space as postulated in the model.  We do not discuss how the model and its parameter space are constructed.  In the usual procedure of model building, a model is generic if its action, Lagrangian or potential includes all renormalizable terms compatible with the presumed symmetries of the theory.  We do not require such an assumption in the following proof.

\section{SUSY vacua from generic R-symmetric models} \label{sec:4}

Consider a continuous family of superpotentials, specified by a set of complex coefficients $\{ c_\alpha \}$ taking values in a continuous parameter space.  The superpotential is constructed from a set of terms $\{ p_\alpha(\phi_i) \}$ allowed by a certain set of conditions including the R-symmetry, and $c_\alpha$ are the coefficients of the linear combination of these terms:
\begin{equation} \label{eq:4-01}
W(\phi_i, c_\alpha) = c_\alpha p_\alpha(\phi_i).
\end{equation}
In a renormalizable Wess-Zumino model, $p_\alpha(\phi_i)$ are monomials of $\{ \phi_i \}$ up to cubic, but here we only need to assume that each term $p_\alpha(\phi_i)$ is at least once differentiable with respect to $\phi_i$ at the vacuum.  Thus $W$ is at least once differentiable with respect to both $\phi_i$ and $c_\alpha$ at the vacuum.

In Section \ref{sec:2}, it is shown that a SUSY vacuum from an R-symmetric Wess-Zumino model always implies a vanishing superpotential.  If we demand that the existence of a SUSY vacuum is generic, the resultant vanishing superpotential is also generic.  Taking $W$ as the function $F$ representing this generic property of SUSY vacua, we have
\begin{equation}
\left. \frac{\ud W}{\ud c_\alpha} \right \lvert_{c_\beta^{(0)}}
    = \left. \left (\frac{\partial W}{\partial c_\alpha}
                    + \frac{\partial W}{\partial \phi_i}
                      \frac{\partial \phi_i}{\partial c_\alpha}
      \right ) \right \lvert_{c_\beta^{(0)}}
    = 0,
\end{equation}
where $c_\alpha^{(0)}$ is a parameter point yielding a SUSY vacuum at $\phi_i(c_\alpha^{(0)})$.  Since $\partial_i W = \frac{\partial W}{\partial \phi_i} = 0$ is satisfied at the SUSY vacuum, our notion of genericity leads to
\begin{equation}
\left. \frac{\partial W}{\partial c_\alpha} \right \lvert_{c_\beta^{(0)}}
    = p_\alpha(\phi_i(c_\beta^{(0)}))
    = 0.
\end{equation}
Thus every term of the superpotential Eq.~\eqref{eq:4-01} vanishes individually at the SUSY vacuum.

The result which we have just proved can be verified by models in literature.  For example, a deformed Polonyi model~\cite{Intriligator:2007cp} has the superpotential
\begin{equation}
W = a X + b X Y^2
\end{equation}
with the R-charge assignment
\begin{equation}
r_{(X)} = 2, \quad
r_{(Y)} = 0.
\end{equation}
It gives two SUSY vacua at
\begin{equation} \label{eq:4-02}
X = 0, \quad
Y = \pm \sqrt{- a / b}.
\end{equation}
The simplest counterexample model~\cite{Sun:2019bnd} has the superpotential
\begin{equation}
W = a X + b X P Q + c X^2 A + d P A^2
\end{equation}
with the R-charge assignment
\begin{equation}
r_{(X)} = 2, \quad
r_{(P)} = 6, \quad
r_{(Q)} = -6, \quad
r_{(A)} = -2.
\end{equation}
It gives a set of SUSY vacua at
\begin{equation} \label{eq:4-03}
X = A = 0, \quad
P Q = - a / b
\end{equation}
with degeneracy on a complex one-dimensional submanifold in the $P$-$Q$ space.  Note that a vacuum at Eq.~\eqref{eq:4-02} keeps the R-symmetry unbroken and a vacuum at Eq.~\eqref{eq:4-03} breaks the R-symmetry with nonzero $a$ and $b$.  One can check that $W$ vanishes term-by-term at the SUSY vacua from both models.  The result proven in this section does not depend on whether the R-symmetry is spontaneously broken at the vacuum or not.

\section{Discussion and Generalizations} \label{sec:5}

In this work, using the formal notion of genericity established in Section \ref{sec:3}, we proved that the expectation value of a generic R-symmetric superpotential vanishes term-by-term at a SUSY vacuum.  This result constrains the form of an R-symmetric Wess-Zumino model which leads to a SUSY vacuum: each term of the superpotential must contain at least one field with a zero expectation value.  Such a constraint may lead to new extensions of the Nelson-Seiberg theorem, contribute to a refined classification of R-symmetric Wess-Zumino models, and find applications in string constructions of $W = 0$ SUSY vacua~\cite{DeWolfe:2004ns, DeWolfe:2005gy, Dine:2005gz, Kanno:2017nub, Kanno:2020kxr} as the first step toward vacua with small superpotentials~\cite{Demirtas:2019sip}.

Although our discussion is based on R-symmetric Wess-Zumino models, the features needed for the proof are just that the superpotential $W$ takes a fixed value at a SUSY vacuum satisfying $\partial_i W = 0$, and that $W(\phi_i, c_\alpha)$ and $\phi_i(c_\alpha)$ are at least once differentiable at the vacuum.  The proof can thus be generalized to any generic function which takes a fixed value at a stationary point.  As an example of such generalizations, consider an action in $d$-dimensional space-time with a scalar potential term
\begin{equation}
S = \int \ud^d x \, V(\phi_i).
\end{equation}
If there is a symmetry which acts non-trivially on space-time coordinates, the volume element $\ud^d x$ may transform as a non-trivial one-dimensional representation under the symmetry.  One example of such a non-trivial symmetry is the scaling symmetry
\begin{equation}
\hat S(\lambda) x = \lambda x, \quad
\lambda \in \mathbb{R}^+.
\end{equation}
The volume element transforms as a charge-$d$ representation, i.e., $\hat S(\lambda) \ud^d x = \lambda^d \ud^d x$.  If the system possesses this scaling symmetry, the potential $V$ must have scaling charge $- d$ to make the action invariant.  A similar argument to the one in Section \ref{sec:2} shows that $V = 0$ holds at a stationary point satisfying $\partial_i V = \frac{\partial V}{\partial \phi_i} = 0$.  Now we assume that $V$ is at least once differentiable with respect to its coefficients $c_\alpha$, which take generic values near $c_\alpha^{(0)}$.  Taking $V$ as the function representing the generic property $V = 0$, the procedure in Section \ref{sec:4} leads to
\begin{equation}
\left. \frac{\partial V}{\partial c_\alpha} \right \lvert_{c_\beta^{(0)}} = 0.
\end{equation}
Thus the expectation of a generic scalar potential with a scaling symmetry vanishes term-by-term at a stationary point.  We expect that this result may find applications in scale-invariant systems such as conformal field theory, phase transitions and critical phenomena, chaos and turbulence, and so on.

\section*{Acknowledgement}

The authors thank Xin Gao, Yan He, Jinmian Li, Peng Li and Bo Ning for helpful discussions.  This work is supported by the National Natural Science Foundation of China under the grant number 11305110.

\end{document}